# Layer-dependent Optical and Dielectric Properties of Large-size PdSe$_2$ Films Grown by Chemical Vapor Deposition

MingYang Wei[a], Jie Lian[a,*], Yu Zhang[a], ChenLin Wang[a], Yueming Wang[a], Zhen Xu[a]

[a] *School of Information Science and Engineering, and Shandong Provincial Key Laboratory of Laser Technology and Application, Shandong University, Qingdao, 250100, Shandong, China*

*: Corresponding author. Email: *jieliansdu@163.com*; lianjie@sdu.edu.cn

## Abstract

Palladium diselenide (PdSe$_2$), a new type of two-dimensional noble metal dihalides (NMDCs), has received widespread attention for its excellent electrical and optoelectronic properties. Herein, high-quality continuous centimeter-scale PdSe$_2$ films with layers in the range of 3L-15L were grown using Chemical Vapor Deposition (CVD) method. The absorption spectra and DFT calculations revealed that the bandgap of the PdSe$_2$ films decreased with increasing number of layers, which is due to the enhancement of orbital hybridization. Spectroscopic ellipsometry (SE) analysis shows that PdSe$_2$ has significant layer-dependent optical and dielectric properties. This is mainly due to the unique strong exciton effect of the thin PdSe$_2$ film in the UV band. In particular, the effect of temperature on the optical properties of PdSe$_2$ films was also observed, and the thermo-optical coefficients of PdSe$_2$ films with different number of layers were calculated. This study provides fundamental guidance for the fabrication and optimization of PdSe$_2$-based optoelectronic devices.

# Introduction

Recently, on account of their unique layer-dependent optical and dielectric properties, two-dimensional (2D) materials have shown a broad application prospect as an essential component of novel optoelectronics devices[1–4]. Currently, a new type of 2D material, noble metal dihalides (NMDCs: MX2, M = Pd, Pt, X = S, Se), has attracted interest due to its apparent layer-dependent physical properties[5–7]. As a representative of NMDCs, palladium diselenide ($PdSe_2$), on account of its unique puckered pentagonal structure[8,9], exhibits strong potential applications in optical, electrical, and thermoelectric aspects[10–15]. Moreover, $PdSe_2$ also has excellent air stability[8,16,17], unique linear dichroism conversion phenomenon[18], high carrier mobility (158 $cm^2/(V·s)$)[8], long-wavelength infrared photo responsivity (~42.1 $AW^{-1}$ at 10.6 μm)[16], and a wide adjustable bandgap (0-1.3 eV, form bulk to mono-layer)[8], making $PdSe_2$ one of the best choice for the next generation of broadband polarization-sensitive photodetectors. At present, few-layered $PdSe_2$ obtained by the mechanical exfoliation method have been manufactured in visible[19,20], infrared[12,21], and near-infrared[16] photodetectors with promising performance. For instance, Zhang et al. [20] fabricated a photodetector based on 5L $PdSe_2$, exhibited competitive capability to detect polarized light with a high photocurrent on /off ratio (higher than 100) and fast response time (less than 11 ms).

Although $PdSe_2$ has shown potential for many applications, the synthesis of large-area continuous $PdSe_2$ films is still a challenge. So far, most of the reported few-layer $PdSe_2$ samples have been achieved by mechanical exfoliation[22–24], molecular beam epitaxy (MBE)[25], and selenization of precursor Pd films[19]. The domain size of $PdSe_2$ prepared by these methods is relatively small (in the micron-scale) and uncontrollable, while the synthesis is inefficient, which seriously limits the application of $PdSe_2$. Chemical vapor deposition (CVD) is a commonly used method to obtain high-quality 2D materials as it enables precise control of the composition[26], thickness[27], and morphology[28] of the samples and has been widely used in the synthesis of $MoS_2$[29], graphene[30], etc. Currently, due to the unique low-symmetry structure of $PdSe_2$, most CVD-grown $PdSe_2$ films are nanosheet in micron-scale[31–33], it is challenging to grow large-area and layer-controlled $PdSe_2$ films. Dielectric function and fundamental optical constants, such as refraction index and extinction coefficient, perform a key role in design and optimization of optoelectronic devices[34]. However, the optical constants and dielectric functions of $PdSe_2$, especially their variation with the number of layers, have not been studied.

Herein, to circumvent this drawback, we report a novel three-zone CVD system that successfully grows centimeter-scale, layer-controlled $PdSe_2$ continuous films on sapphire substrates for the first time. Atomic force microscopy (AFM), X-ray photoelectron spectroscopy (XPS), and Raman spectroscopy analysis show that the grown samples are homogeneous and high-quality. The absorption spectra of the samples were measured using a spectrophotometer, and the optical bandgap of the samples was calculated using the Tauc formula. The results show that the bandgap of $PdSe_2$ films decreases with the increasing number of layers, which is consistent with the first-principle calculation. The refractive index, extinction coefficient and dielectric function of the samples with different layers were obtained by inversion and fitting of the spectroscopic ellipsometer data. The central energy of the exciton peak is obtained using the standard critical point (SCP) model analysis. It is found that the optical constants of $PdSe_2$ have apparent layer dependence due to the interlayer coupling phenomenon. In addition, since optoelectronic devices emit heat in actual use, the effect of temperature on the optical properties of PdSe$_2$ films was also investigated, and the thermo-optical coefficient was calculated. This study paves the way for PdSe$_2$-based wafer-level devices and can provide theoretical guidance for the design and optimization of PdSe$_2$-based optoelectronic devices.

**Result and discussion**

The PdSe$_2$ films were synthesized on sapphire by an innovative three-zone CVD system, as shown in the schematic diagram of Fig. 1a. Centimeter-scale continuous PdSe$_2$ films with a controllable number of layers were grown in the tube furnace using palladium chloride (PdCl$_2$) and selenium (Se) as precursors. The tube furnace is divided into three zones with different temperatures. The Se powder was placed in zone 1 and heated to the melting point of Se (200 °C). The PdCl$_2$ powder was placed in zone 2 while a centimeter-scale sapphire substrate was placed in zone 3 and heated to 300°C and 500 °C, respectively. The evaporated Se and Pd precursors were transported using Ar/H$_2$ at a flow rate of 200/20 sccm as a carrier gas. By controlling the growth time, centimeter-scale PdSe$_2$ films with layers ranging from 3L to 15L were successfully obtained. Fig. 1b presents the optical images of the substrate and the CVD-grown samples. The color changes significantly from light gray to dark gray as the number of layers increases, which is also consistent with absorption spectroscopy. The optical micrograph of the 5L PdSe$_2$ sample is shown in Fig. 1c at a magnification of 100 times. The uniform color contrast of the optical micrograph indicates that the PdSe$_2$ films have excellent thickness uniformity. AFM measurements were used to analyze the morphology and thickness of PdSe$_2$ films, as shown in Fig. 1d and Fig. S1(Supporting information). As PdSe$_2$ film covers the entire surface of the substrate, use pointed tweezers to scratches on the sample to achieve sharp edges before AFM testing. The height difference analysis at the edge of the sample clearly shows that the thicknesses of the prepared PdSe$_2$ films are 1.2 nm, 2.2 nm, 3.3 nm, 5.0 nm, and 6.2 nm, respectively, which are also consistent with the SE measurement (Table S1, Supporting information). Compared with the theoretical PdSe$_2$ crystal layer spacing(0.41 nm)[22,35], the layer numbers of the grown samples can be identified as 3L, 5L, 8L, 12L and 15L, respectively. Moreover, it can be seen from the AFM graphs that the surface of the PdSe$_2$ film (bright areas) is relatively smooth, indicating that the grown PdSe$_2$ film is uniform. Fig. 1e illustrates the schematic illustration of PdSe$_2$ crystal from the top view (up image) and side view (bottom image), clearly expresses the unique pentagonal crystal structure of PdSe$_2$. The chemical composition of the PdSe$_2$ films has been analyzed by XPS. The peaks of Pd 3d and Se 3d are detected from XPS spectrum as indicated in Fig. 1f, where the Pd $3d_{3/2}$ and $3d_{5/2}$ peaks are located at 342.2 eV and 336.9 eV and the Se $3d_{3/2}$ and $3d_{5/2}$ peaks located at 55.8 eV and 54.9 eV, which are consistent with the PdSe$_2$ nanosheets[15] and PdSe$_2$ crystals[19]. More importantly, by calculating the peak intensity and the relative sensitivity factor of XPS[36], the Pd/Se atomic ratio is derived to be close to 2, which is the theoretical stoichiometric ratio of Pd/Se. These results demonstrate that the centimeter-scale PdSe$_2$ films have been synthesized successfully.

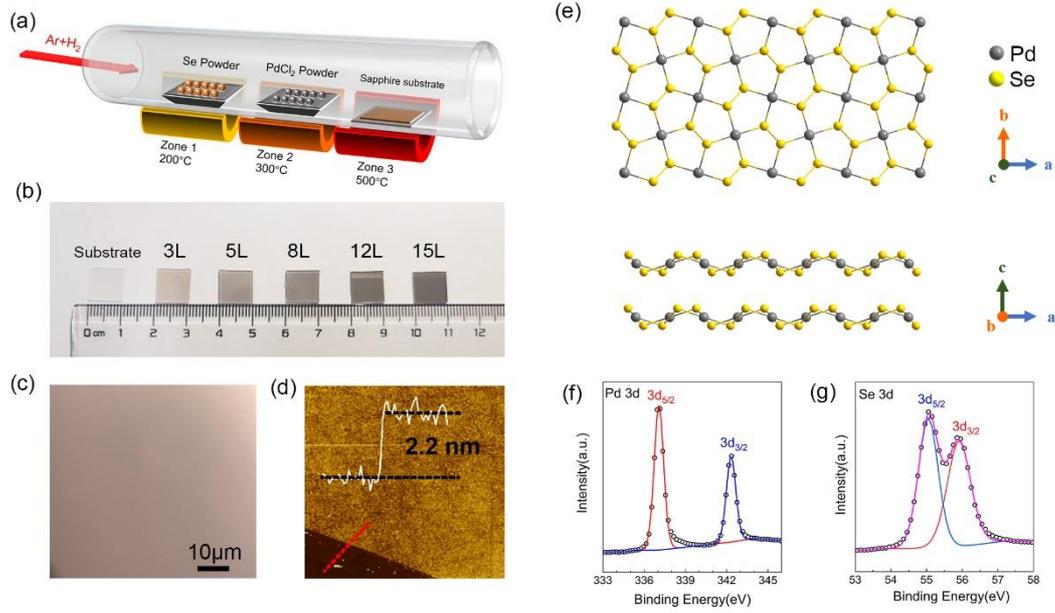

Fig. 1 Synthesis, composition, and morphological characterization of PdSe$_2$ thin films. (a) Schematic diagram of three-zone CVD system. (b) Photographs of PdSe$_2$ with the different number of layers. (c) Optical microscope images of 3L PdSe$_2$ films on a sapphire substrate. (d) AFM image of 5L PdSe$_2$ and the high profiles across the solid white line. (e) The schematic illustration of PdSe$_2$ crystal from top view (up image) and side view (bottom image) (f) XPS spectrum of Pd 3d and Se 3d.

Raman spectroscopy was used to characterize the PdSe$_2$ films with different layers. Previous studies have demonstrated that three A Raman modes ($A_g^1$, $A_g^2$, $A_g^3$) and three B Raman modes ($B_{1g}^1$, $B_{1g}^2$, $B_{1g}^3$) exist in PdSe$_2$ films, which can be ascribed to the in-plane and out of plane modes[18,31]. Noteworthy, owing to $B_{1g}^1$ mode is quite close to $A_g^1$ mode, and its intensity is lower than $A_g^1$ mode[10], only five distinct Raman peaks are shown in Fig. 2a. Raman-inactive modes are activated due to the symmetry reduction in around 100 cm$^{-1}$ to 130 cm$^{-1}$, which did not appear in the Raman spectra of the bulk PdSe$_2$[8]. Except for the peak concerning PdSe$_2$, no other peaks were found within the test range, proving the purity of the growth samples. As illustrated in Fig. 2b, due to the inter-layer coupling, the Raman peaks of PdSe$_2$ redshift with the increasing layer numbers, like the phenomenon that occurs in PtSe$_2$[37] and PtTe$_2$[38]. Specifically, this phenomenon may be attributed to the increased suppression of atomic vibrations by interlayer van der Waals forces as the number of layers increases[39]. These results are consistent with previous research of PdSe$_2$ nanosheets[10,18,32]. To verify the uniformity of centimeter-scale PdSe$_2$ film, Raman mapping scans were performed on 8L PdSe$_2$ film samples tested at 7 mm*7 mm intervals of 1 mm. Fig. 2c(i-iv) shows the intensity of the four Raman peaks ($A_g^1$, $A_g^2$, $B_{1g}^2$, $A_g^3$), the corresponding mapping positions is 146.8 cm$^{-1}$, 208.6 cm$^{-1}$, 224.1 cm$^{-1}$ and 146.8 cm$^{-1}$, respectively. Raman mapping results show that each test point has four Raman peaks, and the intensity at different points of the Raman signal is nearly unchanged, indicates that the PdSe$_2$ we synthesized is uniform over a wide range.

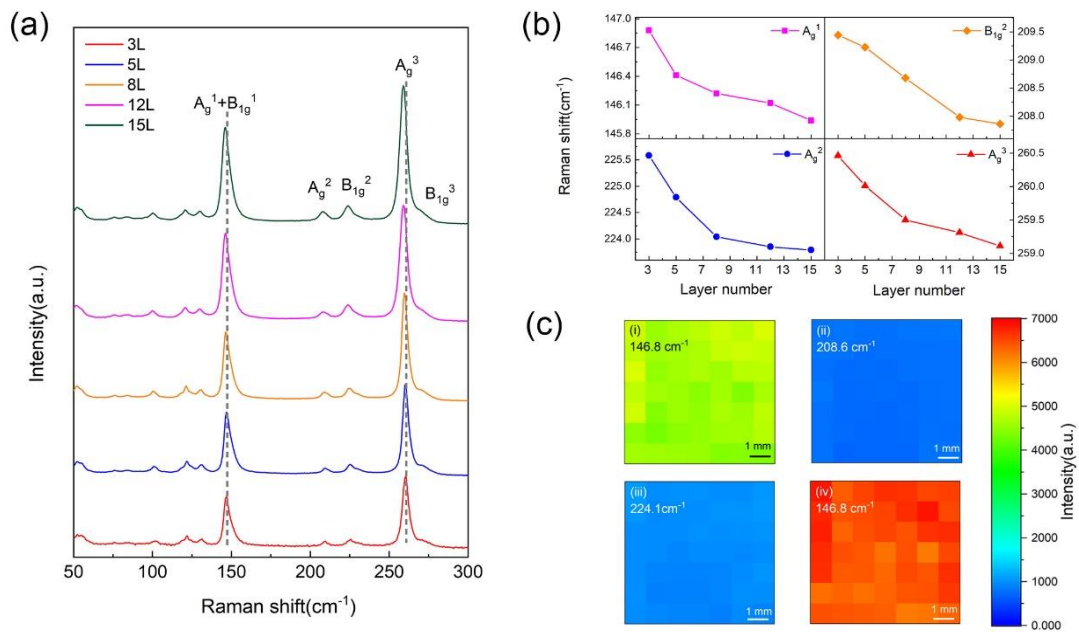

Fig. 2 Raman scattering spectra characterization of centimeter-scale PdSe$_2$ films (a) Raman scattering spectra of PdSe$_2$ films with the different number of layers, the dashed line indicates a redshift of the Raman peak position from 3L to 15L. (b) Raman shifts

value of the four main patterns relative to the number of layers from 3L to 15L. (c) Raman mapping spectrum of 8L PdSe$_2$ films. The mapping range was 7*7mm with 1mm intervals. (i)-(iv) Characterized the Raman intensity at 146.8 cm$^{-1}$, 208.6 cm$^{-1}$, 224.1 cm$^{-1}$ and 146.8 cm$^{-1}$, respectively.

To investigate the optical properties and the layer-related bandgap of PdSe$_2$, the absorption spectra(from 200 nm to 2600 nm) of PdSe$_2$ films with the different number of layers were tested as illustrated in Fig. 3a. Notably, benefiting from the centimeter-scale sample size, the absorption spectrum can be measured directly by a spectrophotometer without the need for a focusing system[15], which makes the testing easier and more accurate. From Fig. 3a, it is easy to see that absorption increases significantly with the increasing layer numbers, which is consistent with optical photographs in Fig 1b. Two distinct absorption peaks, named α and β, can be seen in the absorption spectrum, consistent with the extinction coefficients obtained from SE. The variation of two absorption peak with the number of layers will be discussed in the later sections in conjunction with SE. For layered 2D materials, the absorption spectrum and the Tauc formula are often used to calculate the optical bandgap of the sample[40]. The Tauc formula is as follows:

$$(\alpha h v)^{\rho} = C*(hv - E_g) \qquad (1)$$

where $hv$ and $E_g$ are the incident photon energy and bandgap of the material, and C is a constant. Besides, $\rho$ is the parameter characterizing the optical absorption process, which is theoretically 1/2 and 2 for indirect and direct transition, respectively. Previous studies[8,15] and theoretical calculations have shown that PdSe$_2$ is an indirect bandgap semiconductor, therefore $\rho$ takes the value of 1/2. Consequently, according to equation (1), the bandgap of PdSe$_2$ can be calculated based on the intercept of $(\alpha h\nu)^{1/2}$ and the photon energy map, as illustrated in the inset of Fig. 3a. The results show that the bandgap of PdSe$_2$ films decreases as the number of layers increases, as indicated in Fig. 3b. The band gaps of PdSe$_2$ films of 3L, 5L, 8L, 12L and 15L are 1.05 eV, 0.90 eV, 0.77 eV, 0.67 eV and 0.59 eV, respectively. Moreover, we compared the band gap with the band gap of PdSe$_2$ nanosheet tested by Lu et al.[15]. The results of Lu et al. are consistent with the trend of PdSe$_2$ films; however, our results contain a larger range of layers and better accuracy.

To physically explain the cause of the bandgap reduction phenomenon, we performed theoretical calculations based on DFT on the energy band structure of PdSe$_2$ film, as illustrated in Fig. 3c and Fig. S2(Supporting information). It is worth noting that due to the limitations of DFT calculation in describing long-range many-body interactions, the calculated bandgap is generally smaller than the experimentally obtained bandgap[41]. Nevertheless, the calculations still show an indirect bandgap that decreases as the number of layers increases. The band gaps of the PdSe$_2$ films obtained by DFT are 0.82 eV, 0.68 eV, 0.61 eV, 0.57 eV, and 0.55 eV for 3L, 5L, 8L, 12L, and 15L, respectively. It can be seen from the calculations that the reason for the decrease in the PdSe$_2$ bandgap can be attributed to the orbital hybridization due to the interlayer coupling caused by the increase in the number of layers, as observed in other two-dimensional materials, such as MoS$_2$[42] and black phosphorus[43]. However, since Pd (10 valence electrons) has more valence electrons than both Mo (6 valence electrons) and P (5 valence electrons), it results in stronger orbital hybridization of PdSe$_2$, leading to a greater degree of layer dependent band gaps, giving PdSe$_2$ greater potential for electronic and optoelectronic applications.

The polarization absorption spectrum of the grown PdSe$_2$ film was tested, as shown in Fig. S3(Supplementary Material). The test range of both the non-polarized absorption spectrum and polarized absorption spectrum is 2 mm*2 mm, which is the same as the test range of the ellipsometric spectrometer. The polarization absorption spectrum shows that the absorption of PdSe$_2$ did not change significantly with the angle of polarization. It indicates that the PdSe$_2$ film is isotropic in the tested range. This is related to the fact that the CVD-grown samples are polycrystalline samples so that PdSe$_2$ exhibits isotropy over a wide range.

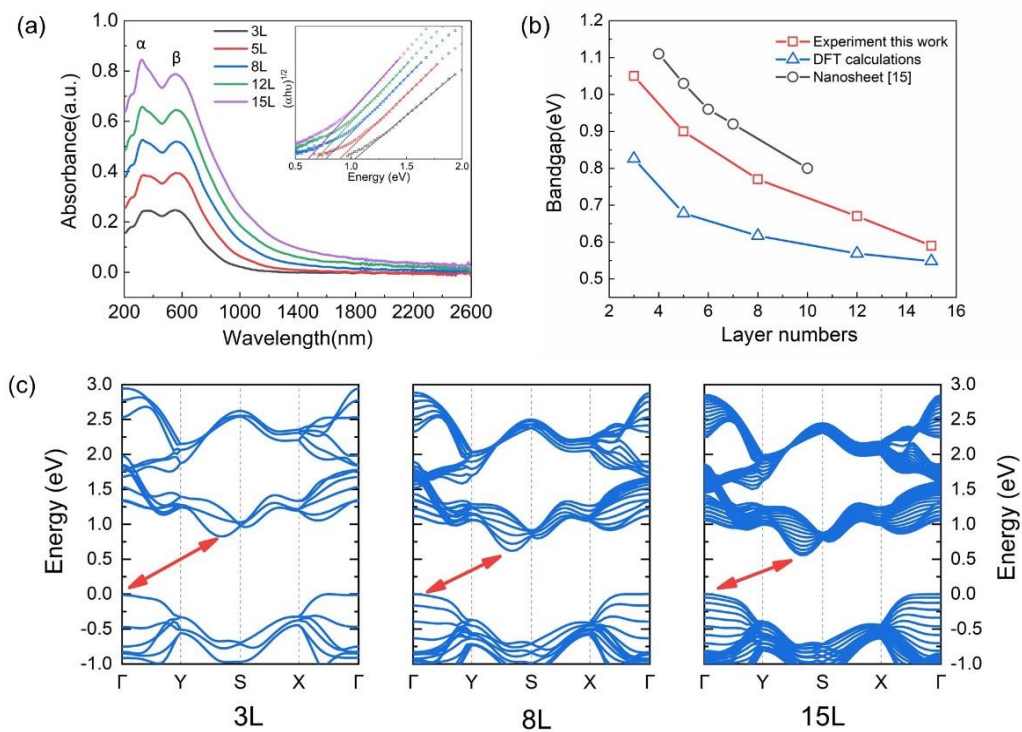

Fig. 3 Layer-dependent bandgap of PdSe$_2$ films. (a) Absorption spectra and the Tauc plot (inset) of PdSe$_2$ films with layer numbers from 3L to 15L. (b) The optical bandgap identified from the Tauc plot (red line) and DFT calculation (blue line) versus layer number. The band gap of the PdSe$_2$ nanosheet is presented for comparison (grey line). (c) DFT calculated energy band structure of PdSe$_2$, the red line shows the indirect bandgap of PdSe$_2$.

In order to obtain the optical constants of PdSe$_2$ and study its variation with the number of layers, different layers of PdSe$_2$ films were characterized by spectroscopic ellipsometry (SE). The absorption spectrum test has shown that the central absorption peak of the PdSe$_2$ films is concentrated at 250-1000 nm, while there is almost no absorption in the near-infrared band. Therefore, we focus on the 245-1000 nm wavelength investigation in the SE research. According to the results of polarized absorption, the sample is isotropic in the test area of SE, and the sapphire substrate has very weak birefringence over the wavelength range tested. Therefore, we used the Jones matrix SE in this work, which is commonly used for isotropic thin-film materials. In spectroscopy ellipsometry analysis, dispersion models are commonly used to describe the dispersion properties of materials in terms of dielectric function and refractive index. In this work, we modeled a three-layer structure (sapphire substrate/PdSe$_2$ film/air) and used the Tauc-Lorentz oscillator model to describe the optical properties of the PdSe$_2$ film. The fitted and measured values of elliptical polarization parameters (Psi and Delta) for the five samples are shown in Fig. S4(Supplementary Material). For all samples, the fitted values are highly consistent with the measured ones. In general, the Mean Square Error (MSE) is used to quantify the deviation of the fitted value from the measured value. It is generally considered that an MSE less than 10 means that the fitting results are reliable. The MSEs of all our samples are less than 2 (see Table S1, Supplementary Material), and the thickness of the sample obtained from SE is very close to that obtained from AFM (see Table S1, Supplementary Material). These all were indicating that the fit results are plausible. More details on spectroscopy ellipsometry analysis can be found in Supplementary Material.

After fitting the measured ellipsometric parameters, the refractive index, extinction coefficient, and dielectric function of the samples can be obtained as shown in Fig. 4. Interestingly, we found that the optical constants and dielectric functions of PdSe$_2$ films are significantly correlated with wavelength and layer numbers. This phenomenon is not apparent in some 2D materials, such as MoS$_2$ and WSe$_2$[44,45]. It is worth noting that the curves of the 12L and 15L samples almost overlap, making the 12L not evident in Fig. 4. In addition, we performed SE tests on PdSe$_2$ films after 6 months of exposure to air, and the results were almost identical to the as-grown films (Fig. S5, Supplementary materials), indicating that the PdSe$_2$ films have good air stability.

The SE analysis showed that the refractive indices of all samples increased sharply with increasing wavelength, exhibiting an anomalous dispersion phenomenon，then slowly decreased after increasing to a specific value. Within 270-600 nm, the refractive index increases with increasing layer numbers. Conversely, the opposite situation is observed within the other wavelengths we tested. Meanwhile, the extensive refractive index of the PdSe$_2$ film demonstrates a large internal scattering cross section[46]. As illustrated in Fig. 4b, the extinction coefficient curve of the PdSe$_2$ film shows an intersection at 465 nm. At 245-465 nm, the smaller the number of layers, the greater the extinction coefficient. The opposite is observed at 465-1000 nm. In fact, the extinction coefficient is related to the absorption of the material[47]. The extinction coefficients obtained by SE corresponded to the α and β peaks in the absorption spectra, and the intersection of the extinction coefficients coincided with the intersection of the α and β peaks in the absorption spectra, providing cross-corroboration of the results. Therefore, combining the SE and absorption spectra measurements, we can analyze that the two absorption peaks of PdSe$_2$ are affected differently by the thickness. This may be related to the alternating dominance of exciton binding energy and complex interlayer interactions. Specifically, for two-dimensional materials, the quantum confinement effect increases as the thickness decreases, which leads to an increase in exciton binding energy[48], resulting in stronger absorption and a larger extinction coefficient. Conversely, complex interlayer interactions lead to stronger absorption in thicker films. Interestingly, previous studies have shown that exciton effects in the high-energy region (UV wavelength region) are more prominent than in the low-energy region[45], which leads to a larger extinction coefficient in the UV wavelength region for the thinner $PdSe_2$ films. Meanwhile, the intersection of the extinction coefficients of the $PdSe_2$ films is almost the same as its unique linear dichroism transition point (472 nm)[18], suggesting that this may have some relationship with linear dichroism transition. The SE analysis also suggests that the relationship between the thickness of $PdSe_2$ and linear dichroism should be further investigated.

The dielectric function of the $PdSe_2$ film versus wavelength is shown in Fig. 4c,d. As described in the supplemental material, since the optical constant and the dielectric function can be transformed into each other, the dielectric function of the $PdSe_2$ film exhibits a similar variation to the optical constant. There is a significant correlation between the value of the $\varepsilon_i$ and the number of layers.

To show this phenomenon more clearly, Fig. 4e extracts the relationship between the value of $\varepsilon_i$ and the number of layers at three commonly used laser wavelengths, which are 364 nm, 532 nm, and 633 nm, respectively. It is noteworthy that the rate of change of $\varepsilon_i$ with thickness of $PdSe_2$ films is relatively small (less than 5%) at 532 nm, showing the potential application utility at this wavelength. In 245-510 nm, the $\varepsilon_i$ at the same wavelength increases with increasing of the layer numbers. However, in 510 nm-1000 nm, the reverse situation appears. We considered that this phenomenon is the result of the alternating domination of exciton binding energy and joint density of states (JDOS). Specifically, according to previous reports[45,49], the value of $\varepsilon_i$ was positively correlated with exciton binding energy and JDOS. As mentioned above, for two-dimensional materials, an increase in the number of layers often leads to a decrease in the exciton binding energy[48,50,51], which results in a decrease in $\varepsilon_i$ as the number of layers increases. JDOS refers to the density of paired initial-final states that participate in the optical leap at a certain energy[45]. While in 2D materials, JDOS shows a positive correlation with the number of layers, and the acceleration decreases gradually[44]. Meanwhile, exciton effects are more prominent in the UV wavelength region than in the IR region [45]. Therefore, we believe that the exciton binding energy is dominant at 245-510 nm, causing $\varepsilon_i$ to decrease with increasing L, while at 510-1000 nm, JDOS plays a dominant role, causing $\varepsilon_i$ to increase with increasing L. Noteworthy, as the thickness increases, the excitonic effect in the 2D material decreases sharply[52], resulting in almost identical optical and dielectric properties for 12L PdSe$_2$ and 15L PdSe$_2$ films.

The exciton peaks in the $\varepsilon_i$ corresponds to the leap between energy bands[53]. As seen in Fig. 4d, two exciton peaks appear in the range of the SE test, labeled A and B. In order to avoid the interaction between exciton peaks and determine the center energy of exciton peaks more accurately, standard critical point (SCP) model was used to fit the second derivative spectrum of dielectric function[54]. The fitting parameters of SCP model can be seen in Table S2 (Supplementary Material). The relationship between the central energies of the A and B exciton peaks and the number of layers is shown in Fig. 4e. It can be seen from the Fig. 4e that for both exciton A and B, their central energies are redshifted as the number of layers increases. This is due to the enhanced interlayer coupling with increasing thickness, leading to significant interlayer orbital hybridization, resulting in a decrease in the central energy of the two exciton peaks. This is in accordance with the variation of the bandgap of PdSe$_2$ films with the number of layers.

As a summary, PdSe$_2$ films has strong layer-dependent optical and dielectric properties when the thickness is less than 12 L, mainly due to the unique strong exciton effect in the UV band, which is not evident in other 2D materials. However, it is not obvious at thicknesses larger than 12 L. This unique property implicates that thin-layer PdSe$_2$ films has excellent prospects for applications in the UV band, such as electro-optical modulators and UV detectors.

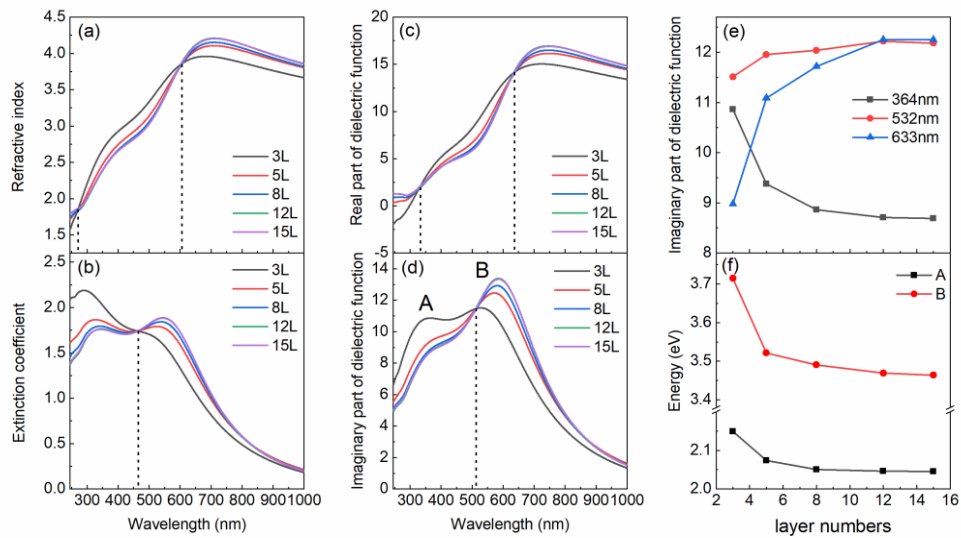

Fig.4 Optical properties of PdSe$_2$ with different layer numbers. (a) Refractive index. (b) Extinction coefficient. (c) The real part of the dielectric function. (d) The imaginary part of the dielectric function. (e) Correlation between $\varepsilon_i$ and layer numbers at three commonly used laser wavelengths. (f) Correlation between the central energy of exciton peaks and layer numbers

In the practical use of optoelectronic devices, high temperature environment is inevitable, to obtain the temperature dependences of optical constants and dielectric function of PdSe$_2$ films, SE measurements were performed at various temperatures (300 K-500 K) and wavelength (between 400 nm to 800 nm). Fig. 5a-d illustrate the refractive index, extinction coefficient, real parts of the dielectric function and imaginary parts of dielectric function versus temperature for 15L PdSe$_2$ films, respectively. As illustrated in Fig. 5a, the refractive index hardly changes with temperature in the wavelength of 400 nm-500 nm, while it decreases with increasing temperature in the wavelength of 500 nm-700 nm. To further investigate the effect of temperature on refractive index, the thermo-optical coefficients of films with the different number of layers were calculated, as indicated in Fig. 5c. The equation for the thermo-optical coefficients is as follows.

$$\xi = dn/dT$$

where $\xi$ represents the thermo-optical coefficient, n represents the refractive index of the sample, T represents thermodynamic temperature. As seen in Fig. 5e, the thermal-optical coefficients of the 8L, 12L and 15L PdSe$_2$ films are around the zero axis in the range of 400 nm-500 nm, reflecting the good thermal-optical stability and potential applications. It is noteworthy that the thermal stability of thin PdSe$_2$ layers is smaller than that of relatively thick PdSe$_2$, which may be attributed to the weaker interlayer coupling in the few-layer PdSe$_2$ films. In the range of 500 nm-700 nm, PdSe$_2$ exhibits a negative thermo-optical coefficient, which may be related to the inherent semi-metallicity in NMDCs[55]. Fig 5c,d shows the temperature dependence of real part of dielectric function and imaginary part of dielectric function. The center energy of exciton peaks at different temperatures were extracted using the SCP model analysis, as indicated in Fig. 5f. A significant red shift of the central energy with increasing temperature can be clearly observed. This could attribute to the enhanced electron-phonon interaction and the expansion of the lattice constant at high temperatures[56–58]. Similar phenomena have been observed from other 2D materials, such as MoS$_2$ and WS$_2$[59]. The study of the temperature-dependent optical properties of PdSe$_2$ films can provide guidance to the practical application of PdSe$_2$-based devices.

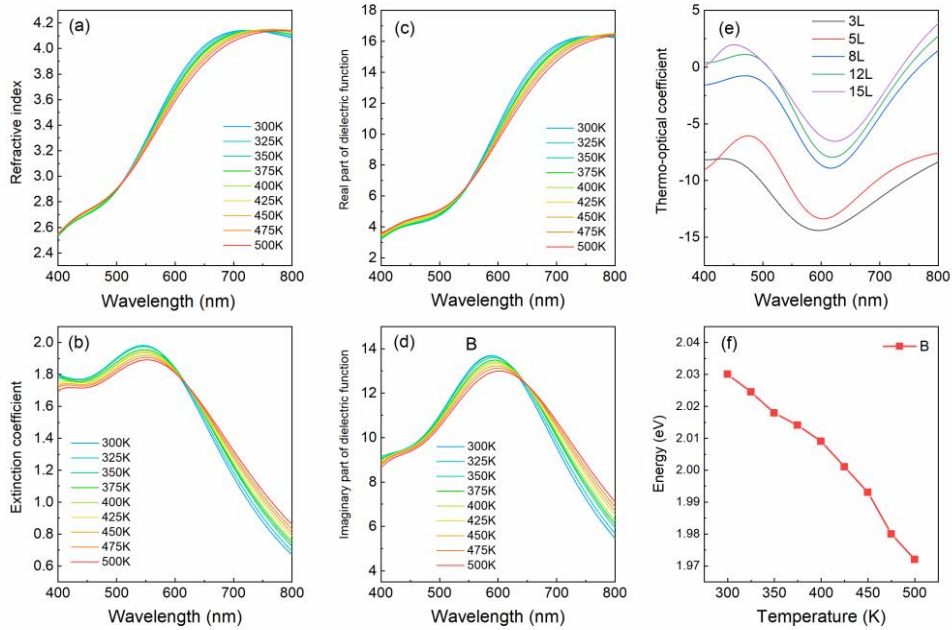

Fig. 5 Temperature-dependent optical and dielectric properties of PdSe$_2$ films. (a)(b)(c)(d) represent the refractive index, extinction coefficient, real part of the dielectric function, and imaginary part of the dielectric function of the 15L PdSe$_2$ film, respectively. (e) Thermo-optical coefficients of PdSe$_2$ films with different number of layers versus wavelength. (f) The central energy of the B exciton of 15L PdSe$_2$ varies with temperature.

## Conclusion

In summary, centimeter-scale continuous PdSe$_2$ films were grown on sapphire substrates using three-zone CVD method with thicknesses from 3L to 5L. The growth quality of the films was verify using Atomic force microscopy, X-ray photoelectron spectroscopy and Raman spectroscopy. As the number of layers increases, the peak positions of the Raman vibration modes are red shifted, which is due to the effect of interlayer coupling. Owing to the enhancement of orbital hybridization, the band gap of PdSe$_2$ films decreases with increasing number of layers, from 1.05 eV at 3L to 0.59 eV at 15L. By inversion and fitting of the spectroscopic ellipsometry (SE) data, the refractive indices, extinction coefficients and dielectric functions of the samples with different layer numbers were obtained. SE analysis shows that PdSe$_2$ has significant layer-dependent optical and dielectric properties. Two exciton peaks are identified in the imaginary part of the dielectric function of PdSe$_2$, and the central energy of the exciton peaks is red-shifted with increasing number of layers. Interestingly, the imaginary part of the dielectric function decreases with increasing layer number from 245 nm to 510 nm, while the opposite result occurs from 510 nm to 1000 nm. We considered that this phenomenon is the result of the alternating domination of exciton binding energy and joint density of states (JDOS). SE analysis also indicates the strong excitonic effect of PdSe$_2$ films in the UV band. Additionally, the effects of temperature on the optical and dielectric properties of PdSe$_2$ were obtained. PdSe$_2$ films exhibits a negative thermo-optical coefficient, which may be related to the presence of semi-metallicity within PdSe$_2$ films. This study provides fundamental information for the design and optimization of PdSe$_2$-based optoelectronics devices, helping one to exploit its potential for a broader range of applications.

**Experimental Section**

Absorption spectra of PdSe$_2$: The absorption spectra of PdSe$_2$ films were measured by double-beam spectrophotometer (UV-3600, Shimadzu). The spectral range of absorption spectrum test is set to 200 nm-3200 nm (0.39 eV-6.2 eV). The PdSe$_2$ films were placed on a sample stage with a light hole, the light area was set at 2mm*2mm to ensure that it matched the test spot range of SE. The polarized absorption spectrum of PdSe$_2$ films was measured using a double-beam spectrophotometer (Cary 5000, Agilent) with a half-wave plate. By keeping the sample stationary and rotating the half-wave plate, the polarized absorption spectrum of the sample was obtained. Both the absorption and polarized absorption spectra were measured at room temperature (300K).

Raman Spectra Measurements: The Raman spectra was tested by LabRAM HR Evolution Raman microscope (HORIBA). The excitation laser wavelength is 532 nm. The spectral resolution is 0.6 nm$^{-1}$ with using an 1800 lines mm$^{-1}$ grating. To reduce testing errors, the acquisition time of the Raman spectra was set to 10 seconds and repeated five times over a range from 50-300 cm$^{-1}$. For Raman mapping scans, the movement of the sample is controlled by a three-dimensional motorized translation table, which performs an automatic focus before each test to avoid focus-induced errors. All Raman tests were performed at room temperature (300 K).

Composition and surface morphology Measurements： The composition measurement was conducted by X-ray Photoelectron Spectrometer (AXIS Supra, Kratos). The XPS data is analyzed and processed by CASAXPS software. The surface morphology of PdSe$_2$ films were obtained by AFM (Smart SPM, AIST-NT) analyzed by the accompanying software.

First-principles calculations: The theoretical calculations of this work are done using the first-principles calculation package Cambridge Sequential Total Energy Package (CASTEP) based on density function theory (DFT)[60]. Exchange-correlation interactions between valence electrons are treated using the Perdew-Burke-Ernzerhof (PBE) under the generalized gradient approximation (GGA). Previous studies have shown that this generalized function allows for reasonable simulations of the electronic and structural properties of PdSe$_2$ layers[8,61,62]. The plane wave energy cutoff is 550 eV to ensure convergence. The Brillouin zone k-space sampling was performed using an 8

× 8 × 1 k-point grid generated by the Monkhorst-Pack scheme. To avoid the effect of interlayer cyclicity, a vacuum layer of 20 Å was set up for each simulation.

Spectroscopic ellipsometry measurement: In this work, a commercial SE with a rotational analyzer (SE-VM, Eoptics Technology) was applied to study the optical and dielectric properties of $PdSe_2$ films. The probing spot diameter of the ellipsometer was 2 mm, in line with the spectrophotometer. The spectral range tested was 245-1000 nm (1.24 eV-5.06 eV), and the spectral resolution was set to 0.6 nm to obtain smoother ellipsometric data. The optical constants of $PdSe_2$ at high temperatures were measured using SE (SE-VE, Eoptics Technology) and an accompanying adjustable temperature sample stage. The measured spectral range was 400 nm-800 nm (1.55 eV-3.1 eV) with a spectral resolution of 1 nm and an incidence angle of 65°. The temperature range of the test was 300 K-500 K, and the heating rate was 5 °C/min, with a minimum temperature control accuracy of 0.1 °C on the sample stage. To minimize the test error, the measured data were obtained after holding for 2 min.

## Acknowledgements

This work is supported by the National Natural Science Foundation of China (grant number 12074214) and the National Key Basic Research Program of China (grant number 2015CB921003).

## Data availability

The data of this study are available from the corresponding author on reasonable request.

## Author Contributions

MingYang Wei assemble the data and write the draft. Jie Lian provides funding acquisition and project administration. Yu Zhang and ChenLin Wang provided theoretical support. Yueming Wang and Zhen Xu performed relevant tests and interpret the results. All authors discussed the results.

## Competing Interests statement

The authors declare no competing interests.